\documentclass[runningheads]{llncs}
\usepackage[width=122mm,left=16.5mm,paperwidth=155mm,height=193mm,top=21mm,paperheight=235mm]{geometry} 
\usepackage[T1]{fontenc}
\usepackage{graphicx,verbatim}
\newcommand{\eg}{e.g., }
\newcommand{\ie}{i.e., }
\usepackage{amsfonts}
\usepackage{amsmath}
\usepackage[capitalise]{cleveref}
\usepackage{booktabs}
\usepackage{multirow}
\usepackage{xcolor}
\usepackage{colortbl}
\usepackage{arydshln}
\usepackage[table]{xcolor}

\begin{document}
\title{DenOiS: Dual-Domain Denoising of Observation and Solution in Ultrasound Image Reconstruction}
\titlerunning{DenOiS: Dual-Domain Denoising of Observation and Solution}
\author{Can Deniz Bezek, Orcun Goksel}
\authorrunning{Bezek \& Goksel}
\institute{Department of Information Technology, Uppsala University, Sweden \\
\email{~}}
\maketitle             
\begin{abstract}
Medical imaging aims to recover underlying tissue properties, using inexact (simplified/linearized) imaging models and often from inaccurate and incomplete measurements.
Analytical reconstruction methods rely on hand-crafted regularization, sensitive to noise assumptions and parameter tuning.
Among deep learning alternatives, plug-and-play (PnP) approaches learn regularization while incorporating imaging physics during inference, outperforming purely data-driven methods. 
The performance of all these approaches, however, still strongly depends on measurement quality and imaging model accuracy.
In this work, we propose \textit{DenOiS}, a framework that denoises both input observations and resulting solution in their respective domains. 
It consists of an observation refinement strategy that corrects degraded measurements while compensating for imaging model simplifications, 
and a diffusion-based PnP reconstruction approach that remains robust under missing measurements. 
DenOiS enables generalization to real data from training only in simulations, resulting in high-fidelity image reconstruction with noisy observations and inexact imaging models.
We demonstrate this for speed-of-sound imaging as a challenging setting of quantitative ultrasound image reconstruction. 
\end{abstract}

\section{Introduction}
Medical imaging aim to recover tissue properties from sensor readings/observations, \eg k-space measurements in magnetic resonance imaging (MRI), sinograms in X-ray computed tomography (CT), and phase shifts in quantitative ultrasound (US). 
These measurements are often incomplete or noisy due to several factors such as limited acquisition time in MRI and US, radiation dose constraints in CT, limited aperture/angle in US and CT, or low signal-to-noise ratio in US. 
Image reconstruction is hence typically ill-posed and solved as an inverse problem with hand-crafted regularization, either directly~\cite{feldkamp_practical_1984} or iteratively~\cite{sauer_local_2002}. 
These methods are sensitive to noise assumptions and initialization, require careful parameter tuning, and commonly based on simplified and linearized physical models.

Deep learning (DL) methods have emerged as powerful alternatives for medical image reconstruction~\cite{ravishankar_image_2019}.
Some methods learn a direct mapping from measurements to images~\cite{zhu_image_2018,chen_robust_2025} (cf.~\cref{fig:method}(a)), which often generalize poorly. 
Other approaches use DL to denoise measurements before applying an analytical reconstruction~\cite{ghani_fast_2019} (cf.~\cref{fig:method}(b)) or to denoise output analytical reconstructions~\cite{jin_deep_2017} (cf.~\cref{fig:method}(c)). 
However, by not integrating the imaging model with DL, such disjoint strategies degrade under distribution shifts and missing measurements~\cite{ravishankar_image_2019}.

Hybrid approaches, instead, aim to incorporate the imaging model into a learned reconstruction process.  
Unrolling-based methods unfold iterative optimization algorithms, \eg the alternating direction method of multipliers (ADMM), into neural network layers to learn parameterized regularization and norms~\cite{sun_deep_2016,adler2018learned,hammernik2018learning,aggarwal_modl_2018,vishnevskiy_deep_2019}. 
Plug-and-play (PnP) methods alternate between a data-consistency step and a pre-trained denoiser~\cite{wu_iterative_2017,zhang_plug_2021} (cf.~\cref{fig:method}(d)). 
Recently diffusion models have shown promise as such denoising priors~\cite{liu_dolce_2023,zhu_denoising_2023}. 
Importantly, unlike unrolling approaches with embedded imaging models, PnP methods with an explicit consistency step allow modifying the model at inference time, \eg for differing undersampling masks in MRI or missing measurements in US.

In this work, we propose Dual-Domain Denoising of Observation and Solution (\textit{DenOiS}), with the following main contributions:
1)~With a derivation showing that measurement (observation) corruption being dependent on the underlying unknown image, we propose a refinement strategy to correct observations, to compensate for simplified/linearized imaging models, to reduce noise, and to impaint missing (undersampled) readings.  
2) We develop a diffusion-model based PnP approach that remains robust under missing measurements. 
3) Combining the two, we propose a reconstruction framework that generalizes well from simulations to real data, demonstrated on in-vivo data with state-of-the-art results.

\section{Methods}
\subsection{Model-based Image Reconstruction and Denoising Diffusion}
A generalized imaging model can be written in discretized form as  
$\mathrm{A}(\mathbf{x}) \mathbf{x} = \mathbf{b}$, 
where $\mathbf{x} \in \mathbb{R}^{N_x N_y}$ denotes the unknown image, 
$\mathbf{b} \in \mathbb{R}^{N_u N_v}$ represents measurements/observations, 
and $\mathrm{A}(\mathbf{x}) \in \mathbb{R}^{N_u N_v \times N_x N_y}$ is the imaging operator based on physics while often depending also on the underlying tissue (\eg scattering in CT and refractions in US) hence leading to a nonlinear problem.  
For practical solutions, most imaging approaches approximate this with a linearized, constant imaging model $\mathrm{A}'$ based on simplified imaging physics; \eg integral operator / Radon transform in CT or Fourier transform (with possible undersampling) in MRI. 
Additionally, in practice, only noisy and potentially incomplete measurements/observations $\mathbf{b}'$ can be made, \eg $\mathbf{b}' = m(\mathbf{b}) + \eta$, 
where $m(\cdot)$ denotes a masking/undersampling operator and $\eta$ represents noise. 
Analytical reconstruction is then posed as a regularized inverse problem:
\begin{equation}
    \label{eq:optimization_problem}
    \mathbf{x}^{\star} = \arg\min_{\mathbf{x}} \| \mathrm{A}'\mathbf{x} - \mathbf{b}' \|_p^p + \lambda \mathrm{\Phi}(\mathbf{x}),
\end{equation}
where $\lambda$ controlling the strength of regularization $\mathrm{\Phi}(\cdot)$. 
An effective strategy for solving \eqref{eq:optimization_problem} is to decouple its data-consistency and regularization terms, which 
half-quadratic splitting does by introducing an auxiliary variable $\mathbf{z}$ and relaxing $\mathbf{x}$$=$$\mathbf{z}$ as a quadratic penalty to solve alternatingly~\cite{venkatakrishnan_plug_2013}:
\begin{equation}
    \label{eq:hqs}
    \mathbf{x}^{\star} = \arg\min_{\mathbf{x}, \mathbf{z}} \| \mathrm{A}'\mathbf{x} - \mathbf{b}' \|_p^p + \lambda \mathrm{\Phi}(\mathbf{z}) + \nu \| \mathbf{x} - \mathbf{z} \|_2^2
\end{equation}
\vspace{-2.3\belowdisplayskip}
\begin{subequations}
\begin{align}
    \label{eq:hqs_a}
    \mathbf{z}_t &= \arg\min_{\mathbf{z}}\; \lambda \mathrm{\Phi}(\mathbf{z}) + \nu_t \| \mathbf{x}_{t-1} - \mathbf{z} \|_2^2 \\
        \label{eq:hqs_b}
    \mathbf{x}_t &= \arg\min_{\mathbf{x}}\; \| \mathrm{A}'\mathbf{x} - \mathbf{b}' \|_p^p + \nu_t \| \mathbf{x} - \mathbf{z}_t \|_2^2 
\end{align}
\end{subequations}
Given the coupling between \eqref{eq:hqs_a} and \eqref{eq:hqs_b}, approximate solutions are often sufficient before iterating.
The \emph{data-consistency} subproblem \eqref{eq:hqs_b} then typically uses a few iterations of a gradient-based method, \eg conjugate-gradient for $p$=2.
In PnP, \eqref{eq:hqs_a} is solved as Gaussian denoising at noise level $\sqrt{\lambda / (2\nu_t)}$. 
Diffusion models address this naturally by sampling from the posterior at the given noise step~\cite{liu_dolce_2023,zhu_denoising_2023}, outperforming CNN-based denoisers~\cite{zhang_beyond_2017,zhang_learning_2017}.

\subsection{Measurement Refinement}
Even with the best computational techniques, solving \eqref{eq:optimization_problem} with inaccurate observations $\mathbf{b}'$ and an inexact model $\mathrm{A}'$ can only find an $\mathbf{x}'$$=$$\mathbf{x}^{\star}$ that satisfies $\mathrm{A}' \mathbf{x}' = \mathbf{b}'$, which can be far from the true solution $\mathbf{x}$.
We aim to find $\mathbf{x}$ with the physical knowledge in $\mathrm{A}'$, by casting approximation errors as learned alterations based on priors identified as follows.
An inexact imaging operator strongly affects any reconstruction~\cite{lunz_learned_2021,bezek_learning_2024}. 
Let $\delta \mathrm{A}(\mathbf{x})$ be the nonlinear error due to simplifying the operator as $\mathrm{A}'$.
Then, the ideal imaging model can be written as:
\begin{equation}
    \label{eq:modified_forwarmodel}
    \mathrm{A}(\mathbf{x})\, \mathbf{x} =  \mathbf{b}
        \quad \Rightarrow \quad 
    \left( \mathrm{A}' + \delta \mathrm{A}(\mathbf{x}) \right) \mathbf{x}
    =  
    \mathbf{b}
    \quad \Rightarrow \quad 
     \mathrm{A}' \mathbf{x} 
     = 
     \mathbf{b} - \delta \mathrm{A}(\mathbf{x})\, \mathbf{x}\,,
\end{equation}
which manifests the need for image-dependent correction to observations. 
To recover ideal $\mathbf{b}$ from $\mathbf{b}'$, one has to \emph{invert} the measurement degradation operator, \eg denoise and inpaint as $\mathbf{b} = m^{-1}(\mathbf{b}' - \eta)$, or in a general form use some \emph{refinement} $f$ as $\mathbf{b} = f(\mathbf{b}')$.
Lacking the knowledge on true image $\mathbf{x}$, we advance by approximating the right side of \eqref{eq:modified_forwarmodel} with its known estimate $\mathbf{x}'$, leading to:
\begin{equation}
    \label{eq:modified_forwarmodel_approximation}
     \mathrm{A}' \mathbf{x} 
     = 
     f(\mathbf{b}') - \delta \mathrm{A}(\mathbf{x})\, \mathbf{x}
     \quad \Rightarrow \quad 
     \mathrm{A}' \mathbf{x} \approx 
     f(\mathbf{b}') - \delta \mathrm{A}(\mathbf{x')}\, \mathbf{x'}
\end{equation}
\subsection{DenOiS: Dual-Domain Denoising of Observations and Solution}
We propose an iterative dual-domain denoising framework that alternates between a measurement refinement network and a DL-based reconstruction module.
Given the above remarks, the refinement process aims to correct non-ideal measurements $\mathbf{b}'$ as a function of noisy/incomplete measurements $\mathbf{b}'$ and an approximate (\eg initial) reconstruction $\mathbf{x}'$.
To make the refinement task domain-consistent, we project $\mathbf{x}'$ into the measurement domain as $\mathrm{A}'\mathbf{x}'$ using its corresponding imaging model.
By also providing any auxiliary prior information $\mathcal{P}$ about the data collection or preprocessing processes (\eg masking/undersampling masks) that may facilitate denoising, we define the 
the refinement network as
${\mathbf{b}} = \mathcal{M}(\mathbf{b}', \mathrm{A}'\mathbf{x}', \mathcal{P})$.
For reconstruction, we use a diffusion-based PnP framework that iteratively denoises a noisy reconstruction $\mathbf{x}_T$ at initial noise temperature $T$ to noise-free $\mathbf{x}_0$.
To help with the updates we provide sample (approximate) precomputed reconstructions $\{\mathbf{x}'_{S}\}$ as conditioning inputs to the diffusion model.
We also provide the measurements $\mathbf{b}$ (when not available, $\mathbf{b}'$ instead) to the PnP framework for data-consistency to be checked regularly via \eqref{eq:hqs_b} alternating with denoising diffusion. 
Denoting these alternating PnP operations by $\mathcal{D}$, reconstruction is performed via $\mathbf{x}_{0} = \mathcal{D}({\mathbf{x}_{T},\mathbf{b}}, \{\mathbf{x}'_S\}).$ 
The resulting $\mathbf{x}_0$ can be re-input to $\mathcal{M}$, iterating the overall scheme.

\begin{figure}[t]
\centering
\includegraphics[width=\textwidth]{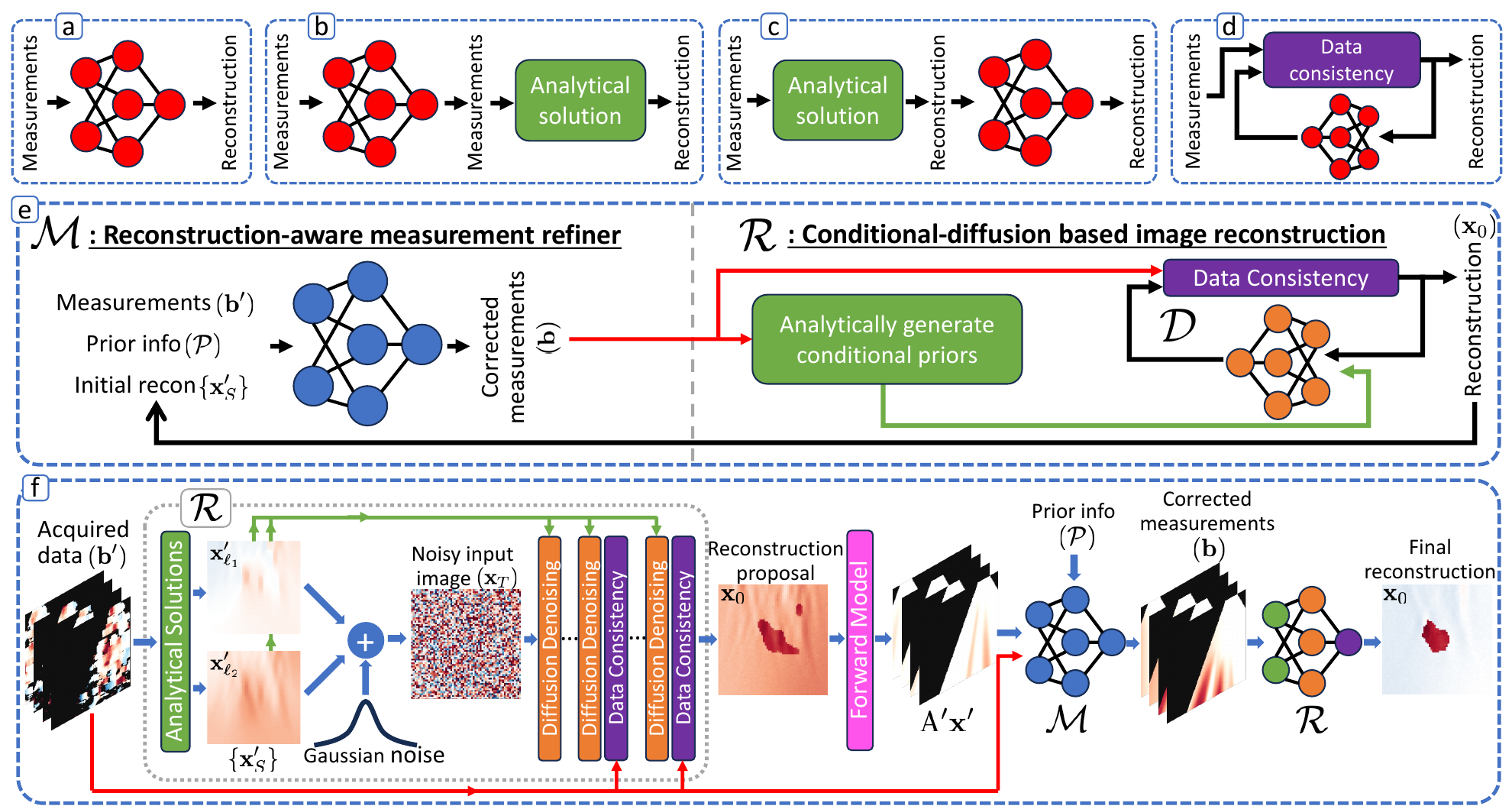}
\caption{Deep learning approaches in image reconstruction (a)~without using the imaging model, (b,c)~using it disjointly, and (d)~jointly with data consistency in a plug-and-play (PnP) framework. (e)~Our proposed iterative dual-domain denoising framework (DenOiS), alternating between observation and solution domain refinement. (f)~Illustration on an ultrasound application for speed-of-sound imaging from incomplete and noisy observations with an inexact model.}
\label{fig:method}
\end{figure}

Conventional diffusion models for image generation from scratch typically start at the highest noise level with pure Gaussian noise. 
In image translation and reconstruction, nevertheless, an initial estimate can be taken as the starting point. 
We use solution proposals $\{\mathbf{x}'_{S}\}$ as good initialization candidates, which improved results in preliminary tests. 
To preserve the noise expectancy of the denoiser, we add Gaussian noise on such initialization to obtain $\mathbf{x}_{T}$ to begin the reverse denoising process. 
We also obtained more robust results by prioritizing denoising with multiple ($K$) consecutive denoising steps before enforcing a data-consistency step. 
The overall framework is illustrated in \cref{fig:method}(e).
\subsection{Application to Quantitative US: Speed-of-Sound Imaging}
We demonstrate our method on quantitative US via imaging speed-of-sound (SoS). 
This is based on an inverse problem relating phase-shifts observed between pairs of transmissions from pulse-echo US to the underlying SoS distribution.
Although a simplified imaging operator $\mathrm{A}'$ is often built via ray-tracing in a limited-angle CT framework~\cite{vishnevskiy_deep_2019}, actual wave propagation $\mathrm{A}(\mathbf{x})$ involves content $\mathbf{x}$ dependent dispersion and refraction as confounders to the assumed wave paths~\cite{li_vivo_2009}.
SoS is an emerging quantitative US biomarker, complementary to B-mode and elastography, for assessing tissue composition / pathology, \eg hepatic steatosis~\cite{imbault_robust_2017}, breast cancer~\cite{schweizer_pulse_2025}, breast density~\cite{bezek_breast_2025}, muscles~\cite{xiao_live_2025}.

To evaluate the proposed method, we performed simulations and acquisitions using a 5\,MHz, 0.3\,mm pitch, 128-element linear transducer with nine linearly-spaced transmits, yielding eight displacement observations $\mathbf{b}'$ for SoS estimation. 
Reconstructions were performed on 64$\times$64 grid with a pixel size of 0.6\,mm. 

\noindent\textbf{Synthetic data (S).}
We analytically generated observations using a simplified ray-tracing based imaging, model based on 10$^4$ synthetically-generated ground-truths $\mathbf{x}$ of quasi-random shapes and SoS distributions in range [1400,1650]\,m/s. 
To prevent bias towards expected inclusions and edges, some $\mathbf{x}$ were relatively homogeneous and 50\% were blurred with random Gaussian kernels.
To emulate discretization errors, we used a twice higher-resolution imaging model $\mathrm{A}_\mathrm{H}$ for the forward simulation $\mathbf{b}$$\leftarrow$$\mathrm{A}_\mathrm{H}\mathbf{x}$, while the reconstructions and learning only had access to a lower-resolution (simplified) version $\mathrm{A}'$ on the intended grid size of 64$\times$64. 
\textbf{Complex in-silico data (C).}
Wave-propagation wise realistic acoustic simulations were conducted on a 37.5\,$\mu$m grid with the k-Wave toolbox~\cite{treeby_kwave_2010}. 
Simulations based on 620 quasi-random shapes as above were employed in training the evaluated models (C$_1$), while additional 30 well-structured circular and rectangular shapes (C$_2$) provided a test set with a distinct distribution.
Both included positive and negative SoS contrast. 
\noindent\textbf{Real data.} Data from four phantom inclusions with manufacturer-reported SoS values as well as in-vivo data from a breast cancer patient with ductal carcinoma were collected. 
\noindent\textbf{Preprocessing.} We beamformed the in-vivo data with 1475\,m/s as the literature-reported average breast SoS, and the in-silico and phantom data with 1510\,m/s.
Phase-shift were estimated using normalized cross-correlation followed by sub-sampling.

\subsection{Implementation}
\noindent\textbf{Reconstruction Module.}
Our reconstruction module, denoted by $\mathcal{R}_\text{D}$, consists of initial solution generation followed by diffusion-based reconstruction. 
We used two conditional inputs $\{\mathbf{x}'_S\}$ as analytical $\ell_1$- and $\ell_2$-based solutions of~\eqref{eq:optimization_problem} with total-variation regularization of relative weight 0.1 for the imaging operator been SVD-normalized. 
For the diffusion model, we use a U-Net backbone~\cite{liu_dolce_2023,dhariwal_diffusion_2021} with four resolution levels with scaling factors (1,2,2,2) and an initial feature-channel size of~64.  
A linear noise schedule ($T$$=$200) is used, and the model is trained with classifier-free guidance (condition dropout rate of 0.2).
The model is pre-trained on dataset S for 400 epochs (learning rate 10\textsuperscript{-3}$\rightarrow$10\textsuperscript{-5}, L2 loss, Adam optimizer) and fine-tuned on C$_1$, to incorporate wave-based errors of the inexact model. 

\noindent\textbf{Measurement Refinement Network ($\mathcal{M}$).}
We test two denoisers: a standard U-Net ($\mathcal{M}_{\text{U}}$) and a diffusion model ($\mathcal{M}_{\text{D}}$).
$\mathcal{M}_{\text{U}}$ uses four resolution levels. 
$\mathcal{M}_{\text{D}}$ follows the $\mathcal{R}_\text{D}$, with an additional level of resolution to accommodate 128$\times$128 measurement data.
Conditional inputs are replaced by the measured displacements and the prior information, which includes (a)~triangular aperture (undersampling) masks indicating the fixed valid measurement areas, (b)~low-correlation masks indicating measurements below tracking SNR, and (c)~directivity masks hinting at angular- and depth-dependent expected variation. 
Treating each image-displacements pair of C$_1$ as a sample (\ie 4960 samples in total), we trained the models for 400 epochs (learning rate $10^{-4}\!\rightarrow\!10^{-6}$, L1 loss, Adam) to convert observed displacements $\mathbf{b}'$ to model-consistent targets $\mathrm{A}'\mathbf{x}$.

\noindent\textbf{Inference $\mathcal{R}_\text{D}$.}
Data consistency is enforced using the conjugate gradient method, with measurements scaled prior to inference to operate directly on the normalized images in the diffusion model. 
For in-silico data, the initial noise level ($T_0$) was set to 15, with data consistency applied every $K$$=$5 denoising steps. 
For real data, we use $T_0$$=$50  and $K$$=$10, due to higher noise level and larger imaging model mismatch. 
Denoising is performed using DDIM sampling~\cite{song_denoising_2020}, which can succeed in few reverse steps when conditionals are informative.\\
\noindent\textbf{Inference $\mathcal{M}_{\text{D}}$.}
With the conditional input being relatively weaker due to missing measurements, we initialized from pure noise ($T_0$$=$200) and used DDPM~\cite{ho_denoising_2020}, which can produce high-fidelity samples given many denoising steps. 
Our implementation on the SoS imaging application is illustrated in \cref{fig:method}(f).

\section{Results}
\noindent\textbf{Metrics:}
\textbf{RMSE} (root mean squared error) when groundtruth is available.
\textbf{SoS contrast} ($\mathrm\Delta \mathrm{c}$) 
as $\tilde{\mu}_{\text{a}}$$-$$\tilde{\mu}_{\text{b}}$ with $\tilde{\mu}$ the median SoS in a tissue region. 
It is clinically relevant, \eg for assessing lesions against potentially varying backgrounds (for evaluation we used the half-cm margin surrounding any lesion). 
\textbf{ACE}~(absolute contrast error)  $|\mathrm\Delta\mathrm{c}_\mathrm{rec}$$-$$\mathrm\Delta\mathrm{c}_\mathrm{gt}|$ assesses error in contrast  when groundtruth is available. 
\textbf{gCNR}~(generalized contrast-to-noise ratio)~\cite{rodriguez_generalized_2019} assesses detectability.

\noindent\textbf{Compared Methods:} We compare with two analytical baselines with TV regularization using an $\ell_2$-norm ($L_2$-TV) and $\ell_1$-norm ($L_1$-TV) as in~\cite{sanabria_spatial_2018,stahli_improved_2020}.
We also compare with two state-of-the-art DL-based methods, for which we used official public implementations training them with our data and pretraining/finetuning approach: 
\textbf{Unet}~\cite{chen_robust_2025} is a recent approach to convert displacements directly to an SoS reconstruction via standard Unet~\cite{ronneberger_unet_2015}. 
To ensure fair comparisons, we performed extensive hyperparameter tuning. 
\textbf{Dolce}~\cite{liu_dolce_2023} is a diffusion-based PnP, that conditions on a single solution, initializes from noise, and alternates each denoising step with data-consistency. 
We conditioned its denoiser on $L_2$-TV reconstruction. 
\textbf{Ablations.} For measurement denoising, we ablate $\mathrm{A}'\mathbf{x}'$ and $\mathcal{P}$ to test contributions from \eqref{eq:modified_forwarmodel_approximation} and observation priors.
To evaluate diffusion for reconstruction, we also compare to a simple Unet as $R$ within our DenOiS framework, denoted as Unet($\mathcal{M}_\text{U}$) trained with corrected measurements.

\begin{figure}[t]
\centering
\includegraphics[width=\linewidth]{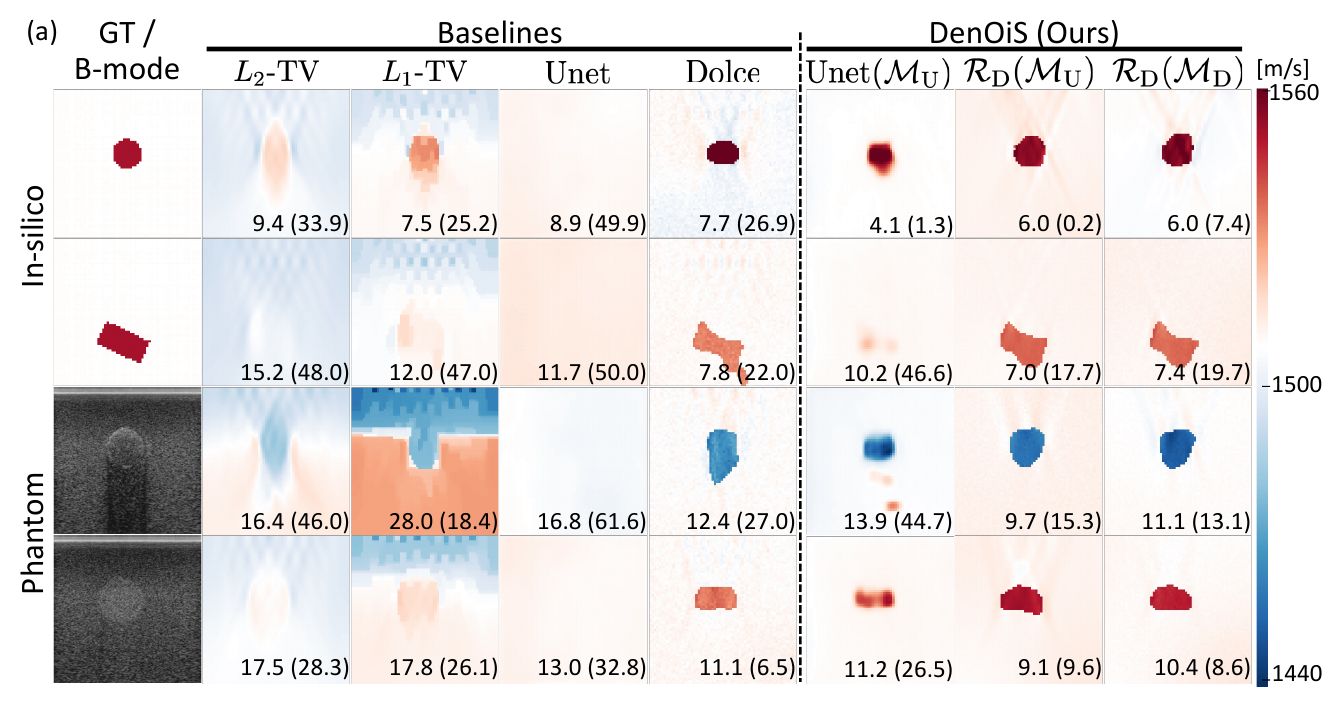}
\includegraphics[width=\linewidth]{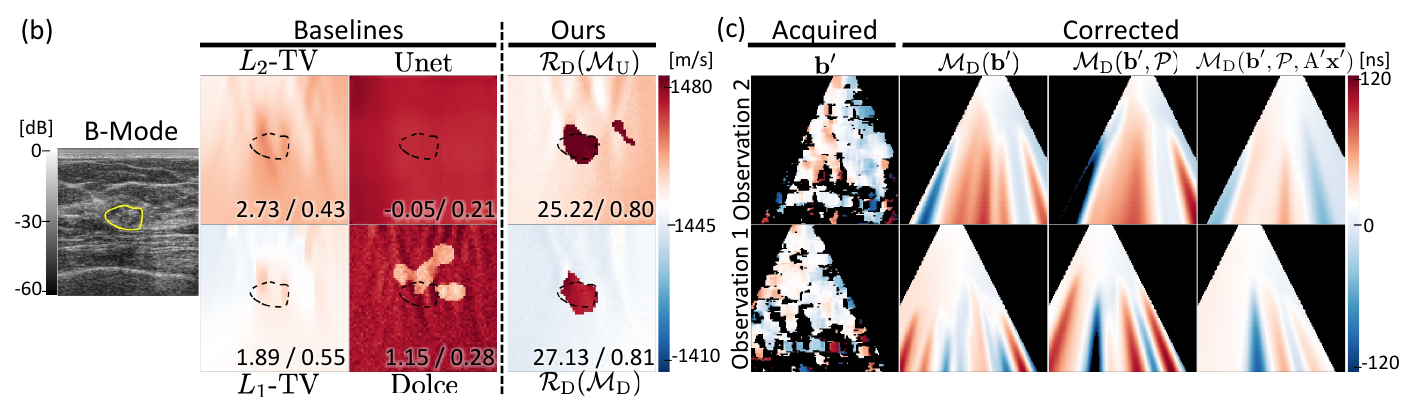}
\caption{(a)~Sample reconstructions for C$_2$ and phantom data, with RMSE (ACE) values reported.
(b)~An in-vivo breast cancer example reporting $\mathrm\Delta \mathrm{c}$ / gCNR. (c)~Corrections of two acquired observations with different $\mathcal{M}_\text{D}$ variants.}
\label{fig:reconstructions_fig}
\end{figure}

\noindent\textbf{In-silico results} (\cref{tab:insilico}). 
\begin{table*}[t]
\centering
\caption{In-silico results on (C$_2$) reporting RMSE (ACE). 
Gray marks baseline configurations.
Rows are observation and columns are reconstruction approaches.}
\label{tab:insilico}
\newcommand{\q}[2]{\textcolor{blue}{#1}&(\textcolor{red}{#2})}
\footnotesize
\setlength{\tabcolsep}{2pt}
\renewcommand{\arraystretch}{1.15}
\resizebox{\textwidth}{!}{
\begin{tabular}{@{}lr@{\,}lr@{\,}lr@{\,}lr@{\,}l|r@{\,}lr@{\,}l@{}}
\toprule
\textbf{Input} 
  & \multicolumn{2}{c}{$L_2$-TV} 
  & \multicolumn{2}{c}{$L_1$-TV}
  & \multicolumn{2}{c}{Unet~\cite{chen_robust_2025}} 
  & \multicolumn{2}{c|}{Dolce~\cite{liu_dolce_2023}}
  & \multicolumn{2}{c}{Unet($\mathcal{M}_\text{U}$)}
  & \multicolumn{2}{c}{$\mathbf{\mathcal{R}_\text{D}}$} \\
\midrule
$\mathbf{b}'$
  & \q{\cellcolor{gray!20}11.95}{\cellcolor{gray!20}36.25}
  & \q{\cellcolor{gray!20}10.39}{\cellcolor{gray!20}32.37}
  & \q{\cellcolor{gray!20}10.73}{\cellcolor{gray!20}42.31}
  & \q{\cellcolor{gray!20}{6.75}}{\cellcolor{gray!20}15.73}
  & \q{43.35}{37.27}
  & \q{6.78}{18.31} \\
    \hdashline
$\mathcal{M}_{\text{U}}(\mathbf{b}')$
  & \q{8.45}{32.23}
  & \q{8.13}{30.43}
  & \q{34.82}{42.27}
  & \q{10.15}{14.46}
  & \q{7.01}{21.22}
  & \q{7.21}{14.35} \\
  $\mathcal{M}_{\text{U}}(\mathbf{b}',\mathcal{P})$
& \q{8.36}{31.39}
  & \q{7.70}{27.35}
  & \q{34.23}{42.27}
  & \q{10.42}{17.56}
  & \q{6.70}{19.38}
  & \q{7.21}{12.28} \\
  $\mathcal{M}_{\text{U}}(\mathbf{b}',\mathcal{P}, \mathrm{A}' \mathbf{x}')$
  & \q{8.26}{30.12}
  & \q{8.04}{28.19}
  & \q{34.20}{42.27}
  & \q{11.78}{27.42}
  & \q{7.10}{24.10}
  & \q{7.55}{11.53} \\
  \hdashline
$\mathcal{M}_{\text{D}}(\mathbf{b}')$
& \q{9.44}{31.04}
  & \q{9.11}{28.87}
  & \q{34.95}{42.27}
  & \q{11.86}{22.65}
  & \q{6.97}{18.04}
  & \q{8.29}{13.16} \\
$\mathcal{M}_{\text{D}}(\mathbf{b}',\mathcal{P})$
  & \q{8.58}{30.32}
  & \q{8.27}{28.23}
  & \q{34.33}{42.27}
  & \q{10.97}{23.62}
  & \q{6.34}{16.58}
  & \q{7.43}{11.67} \\
$\mathcal{M}_{\text{D}}(\mathbf{b}',\mathcal{P}, \mathrm{A}' \mathbf{x}' )$
  & \q{8.24}{29.69}
  & \q{8.13}{28.85}
  & \q{34.37}{42.27}
  & \q{11.89}{28.04}
  & \q{7.23}{24.26}
  & \q{7.59}{12.33} \\
\bottomrule
\end{tabular}
}
\end{table*}
Measurement refinement improves RMSE and ACE for analytical reconstructions. 
Despite numerous training and parameter tuning efforts, the best direct Unet~\cite{chen_robust_2025} model could only find some average SoS value without any inclusion contrast (yielding low RMSE but high ACE), and is excluded from the subsequent tabular results as it degraded even further.
DOLCE achieved low RMSE in-silico, though benefiting little from measurement refinement. 
Our DenOiS framework substantially improves reconstruction quality, with Unet($\mathcal{M}_\text{U}$) achieving low RMSE and ACE and the proposed diffusion reconstruction module $\mathcal{R}_{\text{D}}$ further improving ACE.
\\
\noindent\textbf{Phantom results} (\cref{tab:phantom}). 
\begin{table*}[b]
\centering
\caption{Real phantom data results reporting RMSE (ACE). 
Values within half a standard deviation of the best results 9.79$\pm$5.54 (13.48$\pm$4.0) are highlighted.
Gray marks baseline configurations.}
\label{tab:phantom}
\newcommand{\q}[2]{\textcolor{blue}{#1}&(\textcolor{red}{#2})}
\footnotesize
\setlength{\tabcolsep}{2pt}
\renewcommand{\arraystretch}{1.15}
\resizebox{.9\textwidth}{!}{
\begin{tabular}{@{}lr@{\,}lr@{\,}lr@{\,}l|r@{\,}lr@{\,}l@{}}
\toprule
\textbf{Input} 
  & \multicolumn{2}{c}{$L_2$-TV} 
  & \multicolumn{2}{c}{$L_1$-TV}
  & \multicolumn{2}{c|}{Dolce~\cite{liu_dolce_2023}}
  & \multicolumn{2}{c}{Unet($\mathcal{M}_\text{U}$)}
  & \multicolumn{2}{c}{$\mathbf{\mathcal{R}_\text{D}}$} \\
\midrule
$\mathbf{b}'$
  & \q{\cellcolor{gray!20}18.04}{\cellcolor{gray!20}37.56}
  & \q{\cellcolor{gray!20}21.94}{\cellcolor{gray!20}28.01}
  & \q{\cellcolor{gray!20}14.13}{\cellcolor{gray!20}28.80}
  & \q{34.36}{29.96}
  & \q{13.71}{16.34} \\
    \hdashline
$\mathcal{M}_{\text{U}}(\mathbf{b}')$
  & \q{13.60}{38.41}
  & \q{13.60}{38.25}
  & \q{\cellcolor{yellow!100}10.87}{35.83}
  & \q{14.14}{42.10}
  & \q{\cellcolor{yellow!100}10.26}{24.45} \\
  $\mathcal{M}_{\text{U}}(\mathbf{b}',\mathcal{P})$
& \q{13.12}{38.93}
  & \q{13.18}{38.74}
  & \q{\cellcolor{yellow!100}11.22}{38.25}
  & \q{14.94}{40.60}
  & \q{\cellcolor{yellow!100}10.61}{23.85} \\
  $\mathcal{M}_{\text{U}}(\mathbf{b}',\mathcal{P}, \mathrm{A}' \mathbf{x}' )$
    & \q{13.93}{34.32}
  & \q{13.94}{33.21}
  & \q{\cellcolor{yellow!100}12.11}{31.47}
  & \q{13.49}{30.80}
  & \q{\cellcolor{yellow!100}10.62}{\cellcolor{yellow!100}15.09} \\
  \hdashline
$\mathcal{M}_{\text{D}}(\mathbf{b}')$
& \q{12.95}{36.27}
  & \q{12.88}{35.81}
  & \q{\cellcolor{yellow!100}11.69}{35.18}
  & \q{14.69}{39.52}
  & \q{\cellcolor{yellow!100}10.04}{21.98}  \\
$\mathcal{M}_{\text{D}}(\mathbf{b}',\mathcal{P})$
  & \q{14.01}{34.90}
  & \q{13.99}{34.78}
  & \q{\cellcolor{yellow!100}10.51}{30.31}
  & \q{16.50}{40.03}
  & \q{\cellcolor{yellow!100}\bf 9.79}{20.18}  \\
$\mathcal{M}_{\text{D}}(\mathbf{b}',\mathcal{P}, \mathrm{A}' \mathbf{x}' )$
  & \q{14.78}{32.78}
  & \q{14.76}{32.37}
  & \q{14.19}{30.72}
  & \q{16.94}{30.22}
  & \q{\cellcolor{yellow!100}11.72}{\cellcolor{yellow!100}\bf 13.48} \\
\bottomrule
\end{tabular}
}
\end{table*}
With a large domain shift to realistic data and noise, the results change drastically, to advantage the denoising methods.
Measurement refinement improves results for all, particularly when $\mathrm{A}'\mathbf{x}'$ is provided (see ACE reductions).  
Proposed $\mathcal{R}_{\text{D}}$ outperforms both Dolce~\cite{liu_dolce_2023} and Unet($\mathcal{M}_{\text{U}}$), especially in terms of ACE. 
This highlights the improved generalizability of diffusion-based PnP over direct Unet projection. 
Based on both metrics, we identify two best variants of DenOiS as $\mathcal{R}_{\text{D}}$ with $\mathcal{M}_{\text{U}}(\mathbf{b}',\mathcal{P}, \mathrm{A}' \mathbf{x}')$ and $\mathcal{M}_{\text{D}}(\mathbf{b}',\mathcal{P}, \mathrm{A}' \mathbf{x}')$ abbreviated respectively as $\mathcal{R}_{\text{D}}(\mathcal{M}_{\text{U}})$ and $\mathcal{R}_{\text{D}}(\mathcal{M}_{\text{D}})$ in the sample reconstructions in \cref{fig:reconstructions_fig}(a).
\\
\noindent\textbf{In-vivo Results}
(\cref{fig:reconstructions_fig}). 
DenOiS successfully reconstructs an SoS inclusion at the location matching the clinical annotation, while baseline methods fail on in-vivo data (\cref{fig:reconstructions_fig}(b)). 
While the acquired measurements are noisy and partially missing, our refinement module $\mathcal{M}_{\text{D}}(\mathbf{b}', \mathcal{P}, \mathrm{A}'\mathbf{x}')$ refines them successfully (cf.~\cref{fig:reconstructions_fig}(c)), facilitating the resulting reconstructions. 

\section{Discussion and Conclusion}
Pixel-wise in-vivo groundtruth of tissue properties is unattainable without dicing a subject.
Hence, image reconstruction requires training from in-silico data with two large domain-gaps: physics-wise to real data and complexity-wise to in-vivo heterogeneity.
Learning direct mapping $\mathbf{b}'$$\rightarrow$$\mathbf{x}$ implicitly encodes the training data distribution, thus often failing to generalize, as also seen in our experiments. 
The proposed DenOiS framework enables generalization from in-silico training to real data by jointly denoising observations and reconstructions in their respective domains while also leveraging a physics-based imaging model. 
For measurement refinement, our principled derivation of errors motivated the need for projected reconstructions $\mathrm{A}'\mathbf{x}'$ as a prior, which was validated experimentally as demonstrated in \Cref{tab:insilico,tab:phantom} and \Cref{fig:reconstructions_fig}(c). 
Comparison with state-of-the-art demonstrates the benefits of our proposed multi-conditioning, initialization, and inference strategies. 
DenOiS can be extended to other imaging problems with missing/noisy measurements, \eg MRI undersampling/motion-corruption, CT sparse-view/low-dose/beam-hardening, and PET low photon-count.

\newpage
 \bibliographystyle{splncs04}
 \bibliography{arxiv}

\end{document}